\documentclass[%
reprint,
amsmath,
amssymb,
aps,
]{revtex4-2}
\usepackage[dvipdfmx]{graphicx}
\usepackage{epstopdf}
\usepackage{MnSymbol}
\usepackage{latexsym}
\usepackage{color}
\usepackage{dcolumn}
\usepackage{bm}
\usepackage[mathlines]{lineno}
\usepackage{usebib}

\begin{document}


\title{Detection of Mechanical Deformation Induced by Ultrafast Laser Irradiation upon a Metallic Cantilever}

\author{Takuto Ichikawa}
 \email{s2130052@s.tsukuba.ac.jp}
 \affiliation{Department of Applied Physics, Faculty of Pure and Applied Sciences, University of Tsukuba, 1-1-1 Tennodai, Tsukuba 305-8573, Japan.}%
\author{Aizitiaili Abulikemu}
 \affiliation{Department of Applied Physics, Faculty of Pure and Applied Sciences, University of Tsukuba, 1-1-1 Tennodai, Tsukuba 305-8573, Japan.}
\author{Muneaki Hase}
 \email{mhase@bk.tsukuba.ac.jp}
\affiliation{Department of Applied Physics, Faculty of Pure and Applied Sciences, University of Tsukuba, 1-1-1 Tennodai, Tsukuba 305-8573, Japan.}

\begin{abstract}
In this work, we systematically investigated the ultrafast optical properties of aluminum (Al) thin films on silicon cantilevers using a microscopic femtosecond optical pump-probe technique to explore the effect of light irradiation upon cantilevers while considering radiation pressure and photothermal effects. 
The ultrafast laser pulses used for the study were less than 30~fs in pulse duration and 830~nm in wavelength, and the photon energy (1.49~eV) of the light pulses is close to the interband transition threshold (ITT) of Al.
Therefore, the change in ITT due to the strain of the cantilevers induced by the light irradiation is detected through the change in the transient reflectivity, which is dominated by the thermalized (Th) electron signal. 
We uncovered the position dependency of the transient reflectivity change and the Th electronic signal amplitude of the 100~nm-thick Al films on 160~$\mu$m, 200~$\mu$m and 240~$\mu$m-length cantilevers, and these results are in excellent agreement with two temperature model-based curve fits.
Furthermore, to understand the effect of light irradiation, we derived equations for the position-dependent radiation pressure effect and the photothermal effect, and demonstrated that thermal expansion-induced changes in ITT dominate the position dependence of the signal intensity. Our findings offer avenues for exploring strain effects on ultrafast properties and applications for ultrafast scanning probe microscopy.
\end{abstract}

\date{\today}

\maketitle
\section{\label{sec:level1}introduction}
The irradiation of a material by a pulsed laser is associated with not only elementary excitation such as carriers and phonons but also phenomena such as photothermal  \cite{QIU1992719} and radiation pressure effects  \cite{Ashkin1970}, involving the mechanical deformation of the material. The photothermal effect, a phenomenon where a part of the absorbed light energy is converted into heat, has been discussed as a laser ablation mechanism  \cite{JACQUES1992531} and is the basis of the thermal expansion of the tip in time-resolved scanning tunneling microscopy (STM), or atomic force microscopy (AFM)  \cite{Yoshida2007,Terada2010}. The radiation pressure effect, first proposed by Maxwell \cite{maxwell1873treatise},
is thought to be caused by light imparting momentum to matters. When a photon is reflected or absorbed by a material, a change in momentum occurs, resulting in pressure \cite{Nichols1903,Musen1960}. Sunlight can produce a radiation pressure effect, and its influence on Hayabusa's position control has been discussed  \cite{Ziebart2001,Yamamoto2020}. Moreover, ultrafast laser-induced radiation pressure has been applied as an optical tweezers technique to manipulate nanomaterials and has been attracting increasing attention recently  \cite{Gao2017}.

On the other side, AFM is often used to observe the photothermal and radiation pressure effects. Ma \textit{et al.} separated the photothermal and radiation pressure effects by examining the resonant frequency of the cantilever \cite{Ma2015}. Several other studies have been reported besides them  \cite{ALLEGRINI1992371,Evans2014}, although the experiments are mainly based on existing AFM systems using continuous-wave (CW) lasers. However, observing the ultrafast laser-induced photothermal and radiation pressure effects of the cantilever in more advanced time-resolved STM (or AFM) has rarely been investigated \cite{Yoshida2014}. Therefore, as the first step toward femtosecond time-resolved AFM, we attempted to observe the photothermal and radiation pressure effects of aluminum (Al)-coated silicon cantilevers under femtosecond pulsed laser irradiation.

Optical pump-probe techniques have achieved high temporal resolution using ultrafast pulsed lasers and have been used as experimental methods for investigating ultrafast relaxation dynamics of quasiparticles, such as electron-hole plasma, coherent phonons, and coherent spins in various materials such as nonmagnetic  \cite{PRB11433,PRB024301} and magnetic metals  \cite{PRL4250,Bigot2009}, and semiconductors  \cite{PRB165217,APL1987,APL511987}. The optical excitation from equilibrium to nonequilibrium states is followed by the relaxation processes, during which electron-electron scattering, electron-phonon scattering, and phonon-phonon scattering occur  \cite{KAGANOV1957}. For simple metals, the massive density of free electrons causes rapid damping of electron coherence, thereby dominating the incoherent electron-phonon scattering in the relaxation processes  \cite{PRL1460}. Under such conditions, simplified models for describing thermalization dynamics by considering independently thermalized energy distribution which can be characterized by temperature, in electron and lattice subsystems, are used to compare with experimental data  \cite{PRB12365}. Furthermore, the two-temperature model (TTM) has been used for a coupled system of electrons and lattice for modeling the dynamics in a simple metal  \cite{PhysRevB.71.184301,PRB214305,PhysRevB.68.113102,PhysRevB.91.045129}. Recently, more complex models based on TTM have been produced by considering temperature gradients  \cite{HOHLFELD2000237}, phonon dispersion curve  \cite{PRB100302}, spin system  \cite{PRL166601}, and introducing a nonequilibrium Green's function method   \cite{PRB045128,PRB075126}. However, TTM has not been applied to explore the radiation pressure and photothermal effects on a simple metal.

Al provides a good example of a TTM-based description for electron thermalization because of the fast lattice equilibration by the significant phonon anharmonicity characterized using the large phonon-phonon linewidth  \cite{PRB214305}. These characteristics make Al a suitable medium for femtosecond plasmonic application with surface plasmon polariton waves  \cite{MacDonald2009} and plasmon enhancement for the cantilever tip for the ultrafast scattering-type scanning nearfield optical microscopy (s-SNOM)   \cite{Chen2019}. However, heat accumulation in cantilever tips due to femtosecond pulsed laser irradiation remains a fundamental problem in ultrafast scanning probe microscopy such as ultrafast s-SNOM  \cite{Chen2019}, and the experimental exploration of the interaction between femtosecond pulsed lasers and cantilevers to solve this problem has been lacking.

In this study, we used a thin Al film on a cantilever as a sample of a metallic system where ultrafast dynamics after pulse excitation can be appropriately reproduced using TTM. Ultrashort laser pulses from a mode-locked Ti:sapphire oscillator with a center wavelength of 830~nm (1.49~eV) 
were used as excitation and probe light. The pulse width of optical excitation is shorter than the electron thermalization time constant ($\sim$200~fs), and under ultrafast excitation, we observe transient coherent signals from the sample. 
We observed a remarkable position dependency of the transient reflectivity change ($\Delta R/R$) at an ultrafast timescale. Moreover, we investigated the effects of light irradiation in terms of position-dependent radiation pressure and photothermal properties of Al film on the cantilever and found that the strain mainly originated from the photothermal effect and shows considerable position dependence of the $\Delta R/R$ signal.


\section{\label{sec:level2}EXPERIMENTS}

\begin{figure}
\includegraphics[width=\linewidth]{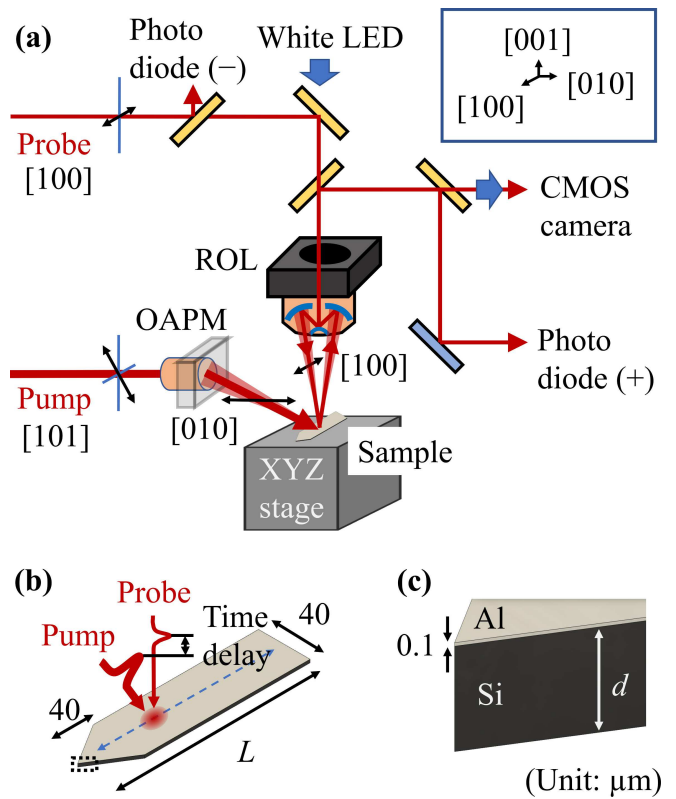}
\caption{(a) Simplified scheme of the optical setup of the microscopic femtosecond optical pump-probe measurement. Each yellow rectangle represents a beam splitter. Each orange cylinder represents a focusing component, which is an off-axis parabolic mirror (OAPM) or a reflective objective lens (ROL). (b) Enlarged view around the sample (the Al-coated Si cantilever) in (a). (c) Enlarged view of the area enclosed by the black dotted line in (b). $L$ and $d$ represent the length and thickness of the cantilever, respectively.}
\label{fig:setup} 
\end{figure}

We performed the reflective pump-probe experiment for the Al film on a cantilever under a microscope field of view using the measurement system shown in Fig.~\ref{fig:setup}(a). The light source used was a mode-locked Ti:sapphire laser with a central wavelength (photon energy $\hbar\omega = 1.49$~eV) of 830~nm, pulse width $\tau_{p}$ of less than 30~fs, and repetition rate $f_{rep}$ of 80~MHz. The laser pulse was split into the strong pump and weak probe pulses using a beam splitter, which traveled through different optical paths. 
The time delay between the pump and probe pulses was scanned at a frequency of 19.5 Hz and an amplitude of 15 ps using an oscillating retroreflector.

All measurements were performed at room temperature $T_{0}$ ($\sim$300~K). Notably, an off-axis parabolic mirror (OAPM) and a reflective objective lens (ROL) were used as focusing components instead of transparent lenses to minimize the dispersion effect and improve the temporal resolution of measurements. Thus, we realized the minimization of the effect of group velocity dispersion of pulse light using reflective components. The pump and probe pulses were focused using OAPM (the depression angle is $\sim$45$^{\circ}$) and ROL with focal lengths of 50.8 and 13.3~mm, respectively. Each shape of the laser spot can be observed in the microscopic image when the sample is opaque and has relatively low reflectivity such as GaAs wafer. Therefore, we observed the Gaussian shape of the focal spot of both pulses on GaAs and obtained the spatial overlapping to detect the signal from the excited carrier using pump pulse irradiation upon the GaAs surface. The pump power was controlled at $\sim$154~mW and the shape of the pump beam was nearly circle by the tilt angle of the kinematic mount for the OAPM. Finally, an oval spot size of pump light with the major and minor axes of 21.1 and 14.4~$\mu$m for the 1/$e^{2}$ width of Gaussian function was demonstrated from the image microscopy, and the corresponding optical fluence was $\sim$800~$\mu$J/cm$^{2}$. Similarly, the size of the probe spot was measured to be 6.6 and 4.2~$\mu$m for the major and minor axes, respectively. The polarization of the focused pump and probe light was orthogonal each other to minimize the effect of the background signal by scattered pump light. Since the direction of the wave vector $\bm{k}$ of the pump light reflected on the OAPM is [$\Bar{1}0\Bar{1}$], [010] can be selected as the direction of polarization. This polarization condition can be realized by the incident [101]-polarized pump light on the OAPM. The energy of the pulse photon is close to the interband transition threshold (ITT)  \cite{PhysRev.147.467,PRB1918,PRB1898}, which enables us to observe a large signal of reflectivity change (divided by reflectivity without optical excitation) $\Delta R/R$. 

Our experimental setup enables us to observe real-time voltage signals related to ultrafast reflectivity changes on a digital oscilloscope after an appropriately amplified and filtered photocurrent signal was detected by a pair of photodiodes, as shown in Fig.~\ref{fig:setup}(a). The photocurrent signal $\Delta I$ equals the current value detected by a photodiode ($-$) $I_{-}$ subtracted from a photodiode ($+$) $I_{+}$. All measurements were performed after maximizing $\Delta I$~$(\propto \Delta R)$ value by monitoring the voltage signal, and normalized reflectivity change $\Delta R/R$ was constructed by the normalization of $\Delta I$ by $I_{+}$~$(\propto R)$ as $\Delta I/I_{+}$~$(=\Delta R/R)$.

Three Al-coated Si cantilevers, whose lengths ($L$) were 160, 200, and 240~$\mu$m, were the samples used [Fig.~\ref{fig:setup}(b)]. 
The respective dimensions (width $w$ and thickness $d$) and mechanical properties (the resonant frequency $f_{0}$ and the spring constant $k_{z}$) of the $L$-length cantilevers are listed in Table~\ref{tab:Alc_properties}.
Al was used as the reflective metal coating to detect the deflection \cite{doi:10.1063/1.100061}.
As illustrated in Fig.~\ref{fig:setup}(c), the thickness of the Al film was estimated to be $\sim$100~nm using a focused ion beam system, whose highest spatial resolution was 5~nm. These cantilevers were carefully attached to the same sample holder to maintain horizontal accuracy, which was confirmed by observing a uniformly focused laser beam on a microscopic image. Each cantilever was attached to the [100] direction, and the sampling position was controlled along the same direction in the measurements.

\begin{table}
\caption{\label{tab:Alc_properties}Length $L$, width $w$, thickness $d$, the resonant frequency $f_{0}$, and the spring constant $k_{z}$ of three Al-coated Si cantilevers used for experiments.}
\begin{ruledtabular}
\begin{tabular}{ccccc}
$L (\mu \mathrm{m})$&$w (\mu \mathrm{m})$&$d (\mu \mathrm{m})$&$f_{0} (\mathrm{kHz})$&$k_{z} (\mathrm{N/m})$\\
\hline
160&40&3.7&$280$&$25$\\
200&40&3.5&$150$&$12$\\
240&40&2.3&$73$&$2.2$\\
\end{tabular}
\end{ruledtabular}
\end{table}

\section{\label{sec:level3}Results and Discussions}
\subsection{\label{subsec:level3a}Fitting based on two-temperature model}

The coupled differential equations using a TTM to describe the dynamics of electron ($T_e$) and lattice ($T_l$) temperatures in Al thin metallic film are given by,

\begin{subequations}
\label{eq:TTM}
\begin{equation}
C_e(T_e)\frac{\partial T_e}{\partial t}=-G(T_e - T_l)+P(t),\label{subeq:TTMa}
\end{equation}
\begin{equation}
C_l\frac{\partial T_l}{\partial t}=G(T_e - T_l),\label{subeq:TTMb}
\end{equation}
\end{subequations}
where $P(t)$ is the absorbed energy density, $C_e(T_e)$ is the heat capacities of the electrons, $C_l$ is the heat capacities of the lattice, and $G$ is the electron-phonon coupling constant \cite{PRB12365, PRB075133}. $C_e(T_e)$ can be approximated by a simple linear dependence on $T_e$ using $C_e=\gamma T_e$ 
in the range of the electron temperature obtained in our experiments ($T_e <$ 2000~K)
 \cite{PRB075133}. 
$\gamma$ is the Sommerfeld constant, which is 135~Jm$^{-3}$K$^{-2}$ in Al \cite{PRB075133}. 
$C_l$ is obtained to be almost constant at 2.30$\times$10$^{6}$~Jm$^{-3}$K$^{-1}$ around room temperature using the Debye temperature (428~K) in Al \cite{PRB075133}. 
In addition, the value of $G$ is nearly constant (2.45$\times$10$^{17}$~Wm$^{-3}$K$^{-1}$) at $T_{0}$ \cite{PRB075133}.
For the expression of $P(t)$, the temporally Gaussian-shaped pulse was commonly used to obtain, 
\begin{equation}
\label{eq:Abs_E}
P(t)=\frac{(1-R)F_p}{\delta\tau_p}exp\left(-\frac{t^2}{\tau_p^2}\right),
\end{equation}
where $F_p$ is the pump fluence, $\delta$ is the optical penetration depth, and $\tau_p$ is the laser pulse width. $\delta = 9.65$~nm at the central wavelength of 830~nm  \cite{Cheng2016} is used for simulating our experimental condition. Using $F_p=800$~$\mu$J/cm$^{2}$ and $\tau_p =30$~fs to numerically solve Eqs.~(\ref{subeq:TTMa}) and (\ref{subeq:TTMb}), we obtain the maximum value of $\Delta T_e\sim$1700~K. For $T_e <$ 2000~K, the assumption of $C_e=\gamma T_e$ is valid, and the time scale of electrons and lattice thermalization is comparable to the experimental results ($\sim$1.0~ps). However, the observed initial coherent signal cannot be reproduced by Eqs.~(\ref{subeq:TTMa}) and (\ref{subeq:TTMb}) (data not shown), suggesting that an additional term related to nonthermal (NT) electrons must reproduce faster relaxation signals. Sun \textit{et al.} \cite{PRB12365} introduced the NT terms and analytical solutions were derived when $C_e\ll C_l$ and $C_e$ and $G$ are constants,
\begin{subequations}
\label{eq:TTM3}
\begin{equation}
N \propto H(t) \exp \left(-\frac{t}{\tau_{NT}}\right) ,\label{subeq:TTM3a}
\end{equation}
\begin{equation}
C_e \Delta T_e \propto \Delta T_e \propto H(t) \left[1-\exp\left(-\frac{t}{\tau_{Th}}\right)\right]\exp \left(-\frac{t}{\tau_{ep}}\right) ,\label{subeq:TTM3b}
\end{equation}
\end{subequations}
where $N$ is the energy density stored in the NT electron distribution, $H(t)$ is the Heaviside step function, $\tau_{NT}$ is the decay time of the NT electron population, and $\tau_{Th}$ and $\tau_{ep}$ are the rise and decay times of the thermalized electron population, respectively \cite{PRB12365}. 

\begin{figure}
\includegraphics[width=\linewidth]{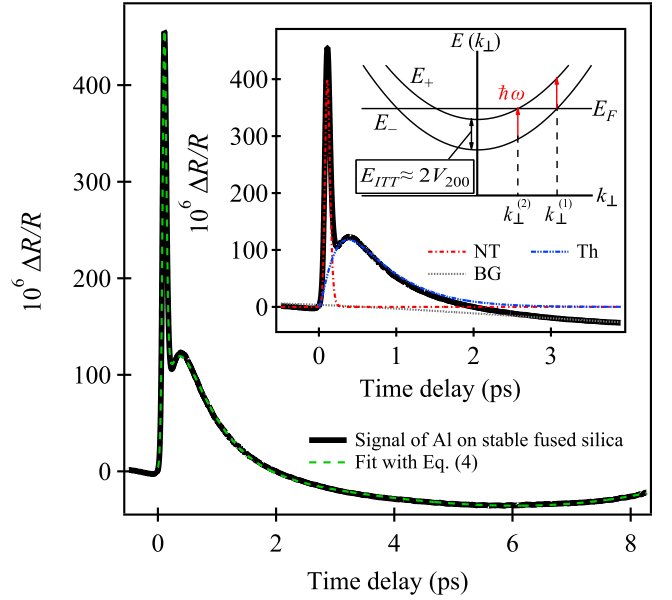}
\caption{The transient reflectivity change $\Delta R/R$ signal of the 100 nm-thick Al film on stable fused silica. The thick black line represents the raw data, and the green dashed lines represent a curve fit using Eq.~(\ref{eq:fit_func}). Fitting components of NT (the first term of Eq.~(\ref{eq:fit_func})), Th (the second term of Eq.~(\ref{eq:fit_func})), and BG (the third term of Eq.~(\ref{eq:fit_func})) are shown in red, blue, and gray-dashed lines, respectively, overlapped onto the measured signal (thick black line) in the inset. The inset shows a schematic of the energy bands (the lower band is $E_-$ and the higher band is $E_+$) mapped in a plane to the (200) face of the Brillouin zone in Al, based on Ref.~\cite{PRB1898}. The horizontal black line represents the Fermi level, while the vertical red arrows indicate interband transition by the photon with the energy of $\hbar\omega > E_{ITT}$.}
\label{fig:rAlm} 
\end{figure}

Since $C_e (= \gamma T_e)$ is no longer constant for a large change in $T_e$ from 300~K to 2000~K, we introduced a stretch coefficient $\alpha_s$ into Eq.~(\ref{subeq:TTM3b}) considering $T_e$ dependency of $C_e$ \cite{PRB075133}. Finally, introducing error functions as the convolution of $H(t)$ and Gaussian function with a width of $\tau_p$ and a background (BG), the fitting function for our experiments can be expressed as,
\begin{eqnarray}
\frac{\Delta R(t)}{R}=&&\frac{A_{NT}}{2}\left[\mathrm{erf}\left( \frac{t-t_0}{\tau_p} \right)+1 \right] \exp\left(-\frac{t-t_0}{\tau_{NT}}\right)\nonumber\\
&&+\frac{A_{Th}}{2}\left[\mathrm{erf}\left( \frac{t}{\tau_p} \right)+1 \right]\nonumber\\
&&\times\left\{1-\exp\left[-\left(\frac{t}{\tau_{Th}}\right)^{\alpha_s}\right]\right\}\exp \left[-\left(\frac{t}{\tau_{ep}}\right)^{\alpha_s}\right]\nonumber\\
&&+ O(t^3),\label{eq:fit_func}
\end{eqnarray}
where $A_{NT}$ and $A_{Th}$ are the NT and Th electronic signal amplitudes, respectively, and $O(t^3)$ is the BG term, assuming a cubic function. 
We confirmed the validity of the fitting function by applying it to the measured $\Delta R(t)/R$ signal of 100 nm-thick Al film on stable fused silica, as shown in Fig.~\ref{fig:rAlm}. 
The fit result is satisfactory and we obtain the parameters of $A_{NT}= (711\pm3)\times10^{-6}$, $\tau_{p} = 38.3\pm0.1$~fs, $\tau_{NT} = 22.8\pm0.1$~fs, $A_{Th} = (195\pm3)\times10^{-6}$, $\tau_{Th} = 234\pm2$~fs, $\tau_{ep} = 860\pm8$~fs, $t_{0} = 125$~fs, and $\alpha_{s} = 1.32\pm0.01$. The same fitting was possible for the Al film on the cantilever shown in the next section.
The photoabsorption originating from the parallel band structure of Al, as reported in Ref.~\cite{PRB1898}, dominates the $\Delta R/R$ signal in the recent femtosecond optical pump-probe experiments \cite{PhysRevB.71.184301,HOHLFELD2000237,PhysRevLett.58.1680}, that is also the case in the present study, as discussed below.

The effective crystal potential $V_{200}$ parallel to the (200) plane in Al solves the degeneracy of the bands.
The vertical electronic transition from the lower ($E_-$) to the higher bands ($E_+$) over the Fermi level occurs 
for the wavevector between $k_{\perp}^{(1)}$ and $k_{\perp}^{(2)}$, as indicated in the inset of Fig.~\ref{fig:rAlm}  \cite{PhysRev.147.467,PRB1918,PRB1898}.
The interband transition threshold $E_{ITT}$ is approximately equal to 2$V_{200}$, and the contribution of this transition to the imaginary part of the dielectric constant $\epsilon_{2}$ is given by 
$\epsilon_{2}(\hbar\omega, E_{ITT}) \propto (\hbar\omega)^{-1}(\hbar\omega - E_{ITT})^{-1/2}$ for $\hbar\omega > E_{ITT}$ \cite{PhysRev.147.467,PRB1898,PhysRevB.24.892}.
In the Th electron system, since the density of electronic states in the vicinity of the Fermi level can be described by the Fermi distribution with an electron temperature $T_{e}$, and $\Delta R/R$ signal is dominated by the contribution of $\Delta \epsilon_2$ due to the change in the electron temperature, $\Delta T_{e}$, the Th electronic signal amplitudes $A_{Th}$ is proportional to the change in $E_{ITT}$, 
i.e., $A_{Th}$ $\propto$ $\Delta T_{e}$ $\propto$ $\Delta \epsilon_{2}$ $\propto$ $\Delta E_{ITT}$.

\subsection{\label{subsec:level3b}Position dependent time-domain signals}

\begin{figure}
\includegraphics[width=\linewidth]{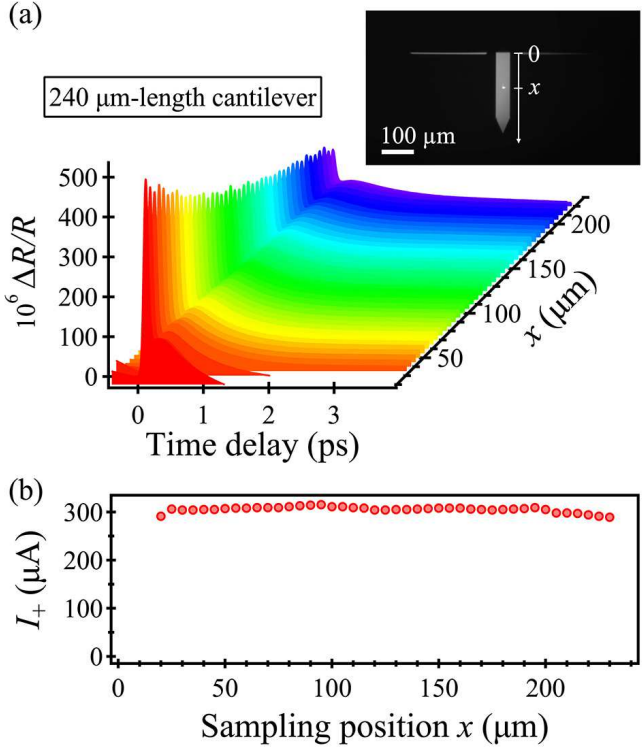}
\caption{\label{fig:rAlc240}
(a) The position-dependent time-domain $\Delta R/R$ signals of the 100 nm-thick Al film on the 240 $\mu$m-length cantilevers. The sampling position $x$ ($\mathrm{\mu m}$) is expressed by the distance from the root of the cantilever. Forty-three signals were aligned from $x=20$ (front) to $x=230$ (back). The smallest (largest) value of $x$ was determined by the limit of the accurate measurement because of the scattering of the pump spot from the root (end) of the cantilever. (b) The position-dependent current value of $I_{+} $ used for the normalization of $\Delta I$~$(\propto \Delta R)$.}
\end{figure}

The position-dependent $\Delta R/R$ signals of the 100 nm-thick Al film on the 240~$\mu$m-length cantilevers are shown in Fig.~\ref{fig:rAlc240}(a), with the probe photocurrent of $I_{+}(\propto R)$ in Fig.~\ref{fig:rAlc240}(b). 
The transient reflectivity change signals near the root of the cantilever ($\Delta R/R\sim5\times10^{-4}$ for $x=20$~$\mu$m) is the same level of Al film on a stable substrate shown in Fig.~\ref{fig:rAlm}. The $\Delta R/R$ signal gradually decreases as the sampling position shifts from the root to the endpoint on a cantilever [see Fig.~\ref{fig:rAlc240}(a)]. 
Here, we performed the same measurements for the 200~$\mu$m- and 160~$\mu$m-length cantilevers, and confirmed the similar tendencies, indicating the phenomena observed in Fig.~\ref{fig:rAlc240} were common.
This means that a dramatic decrease in the $\Delta R/R$ signal cannot be explained by the change in the static reflectivity $R$ at different positions as discussed below.

\begin{figure}
\includegraphics[width=\linewidth]{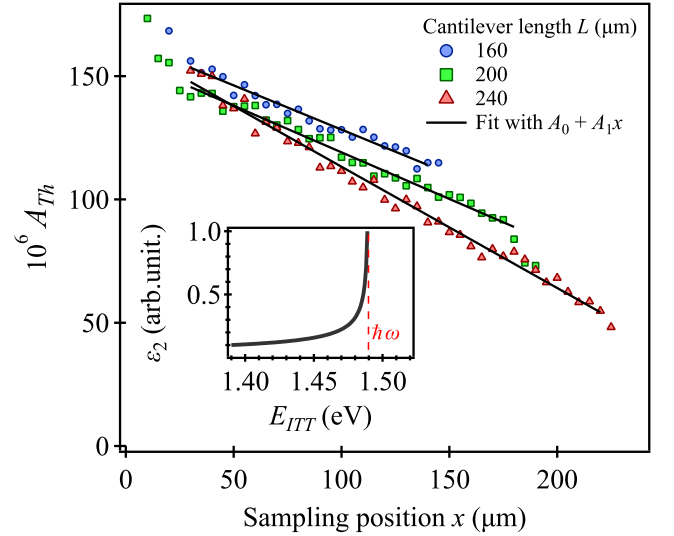}
\caption{\label{fig:rAlc_coef} 
The position-dependence of the thermalized electronic signal amplitude $A_{Th}$ obtained from 100-nm-thick Al films on 160~$\mu$m, 200~$\mu$m, and 240~$\mu$m-length cantilevers are shown with a blue circle, green square, and red triangle markers, respectively. The black lines represent linear function fits ($A_0+A_1x$) in the range of $30\leq x \leq L-20$. The inset presents the imaginary part of the dielectric constant $\epsilon_{2}(\hbar\omega, E_{ITT}) \propto (\hbar\omega)^{-1}(\hbar\omega - E_{ITT})^{-1/2}$ with $\hbar\omega=1.49$~eV as a function of $E_{ITT}$.}
\end{figure}

Each signal was fitted by using Eq.~(\ref{eq:fit_func}), and the results of position dependency of fitting coefficients $A_{Th}$ are shown in Fig.~\ref{fig:rAlc_coef}. 
All parameters were successfully extracted using the curve fitting.
In Fig.~\ref{fig:rAlc_coef}, we found a clear position dependency reflected on the Th electronic signal amplitude $A_{Th}$ and the same dependency is observed in all three cantilevers. 
This is attributed to the position dependency of $\Delta R$ because the values of $I_{+}$ ($\propto R)$ obtained at different sampling positions are nearly constant (fluctuation was less than $10\%$) as shown in Fig.~\ref{fig:rAlc240}(b). 
To compare the rate of decrease of $A_{Th}$, we fit the data using a linear function of $A_0+A_1x$, where $A_0$ is the constant for $x=0$ and $A_1$ is the variation with respect to $x$. 
Furthermore, we selected the fitting range $30\leq x \leq L-20$ since the experimental errors near the root of the cantilever largely prevent obtaining the precise values for $A_{0}$ and $A_{1}$. 
The same fitting was also performed for other parameters ($A_{NT}$, $\tau_p$, $\tau_{NT}$, $\tau_{Th}$ and $\tau_{ep}$) and the results are shown in Appendix~\ref{sec:Ap_A}.
Although $t_{0}$ and $\alpha_{s}$ did not show a linear trend, the values ($t_{0}$:~125 - 150~fs, $\alpha_{s}$:~1.4 - 1.5) were close to the fit results for Al film on stable fused silica (data not shown).
It is notable that the electron thermalization time constant $\tau_{Th}$ ($\sim$200 fs) obtained from the cantilevers is sufficiently longer than the pulse length $\tau_{p}$ ($\sim$30 fs), allowing for the separation of NT and Th electronic signals.

Next, to clarify the mechanism of the linear position dependence of the Th electron signal amplitude $A_{Th}(\propto\Delta E_{ITT})$, we investigate the main position-dependent effects on the cantilever: the radiation pressure effect and the photothermal effect \cite{Ma2015}. However, under our experimental conditions, the pump and probe beams are irradiated at the same position at 80 MHz repetition rates, the radiation pressure effect is found to be negligible by the reasons below.  First, the magnitude of the bending moment, and the proportional strain, are zero at the irradiated position, which is discussed in more detail in Appendix~\ref{sec:Ap_B}. In Appendix~\ref{sec:Ap_B}, we also obtained an expression for the vertical deviation due to light pressure, which indicates that the magnitude of the deviation has an irradiation position dependence. However, since the strain induced by the cantilever occurs after a relatively long time delay, the response for radiation pressure is time averaged and is found to be only 0.3~nm for maximum deviation. Second, at the pulse duration (30~fs), the longitudinal force induced by the radiation pressure is 274~$\mu$N, which results in compression of the material. However, since the deformation area estimated from the characteristic longitudinal velocity of 6430~m/s~\cite{Grossmann_2017} is only 0.2~nm, no strain due to the bending of the cantilever occurs within the pulse duration (30 fs).

On the other hand, considering the effect of thermal expansion of the crystal lattice by temperature rise as a photothermal effect, we found that a significant change in $E_{ITT}$ is expected. If the length $L$ of the cantilever is sufficiently larger than its width $w$ and thickness $d$, the lattice temperature $T_l$ is assumed to be uniform over the entire cross-section, and the temperature field $T_l(x)$ for position $x$ can be described by the one-dimensional thermal diffusion equation \cite{WOS:000649074000009}.
In our experiments, the temperature rise $\Delta T_l(x)$ can be evaluated by the following equation for bi-layered cantilevers \cite{doi:10.1063/1.4795625,doi:10.1063/1.1144509,doi:10.1063/1.2829999,s7091757},
\begin{equation}
\label{eq:1D_dif_eq_sol3}
\Delta T_{l}(x)=\frac{(1-R)P}{w(\lambda_{\mathrm{Al}}d_{\mathrm{Al}}+\lambda_{\mathrm{Si}}d_{\mathrm{Si}})}x,
\end{equation}
where $d_{\mathrm{Al}}=0.1$~$\mu$m and $d_{\mathrm{Si}}(=d-d_{\mathrm{Al}})$ are the thicknesses of Al and Si, respectively, and $\lambda_{\mathrm{Al}}$ and $\lambda_{\mathrm{Si}}$ are the thermal conductivities of Al and Si, respectively \cite{doi:10.1063/1.4795625,doi:10.1063/1.1144509}.
At a temperature of 300~K, $\lambda_{\mathrm{Al}} =$~237 Wm$^{-1}$K$^{-1}$ \cite{White1997} and $\lambda_{\mathrm{Si}} =$~156 Wm$^{-1}$K$^{-1}$ \cite{PhysRev.134.A1058}. 
From Eq.~(\ref{eq:1D_dif_eq_sol3}), the maximum of $\Delta T_{l}$ realized in this experiment is derived to be $\sim$474 K when irradiated to the free end ($x=L$) of a 240~$\mu$m cantilever. It is notable that the linear position $x$ dependence of $A_{Th}(x)$ observed in our experimental results and the relationship between cantilever thickness $d$ and the rate of decrease of $A_{Th}(x)$ with respect to position $x$ can also be explained by using Eq.~(\ref{eq:1D_dif_eq_sol3}).


We estimate the change in $E_{ITT}$ resulting from the thermal expansion in this temperature rise $\Delta T_{l}$. Using the coefficient of thermal expansion of Al at 300~K of $2.33\times 10^{-5}$/K \cite{doi:10.1080/01418610008212140}, the strain yields $1.1\times 10^{-2}$. The relationship between the strain and $\Delta E_{ITT}$ of the Al film can be expressed using the deformation potential \cite{Yu1996} $dV_{200}/de$ as
\begin{equation}
\label{eq:E_ITT_strain}
\Delta E_{ITT}(e)=2\Delta V_{200}(e)=2(dV_{200}/de)e.
\end{equation}
The value of the deformation potential $dV_{200}/de$ depends on the axiality of the strain \cite{JILES1984327}. Since thermal expansion is isotropic, $-3.59$~eV was used as the value of $dV_{200}/de$ \cite{JILES1984327}. Applying the above distortion to the equation, the resulting $\Delta E_{ITT}$ is $-79$~meV. This means that when the initial $E_{ITT}$ is $\sim$1.49 eV, it decreases to $\sim$1.41 eV due to the temperature increase, and the inset of Fig.~\ref{fig:rAlc_coef} shows that the value of $\epsilon_{2}$ can be reduced by $\sim$89$\%$ due to $\Delta E_{ITT}$. Therefore, the large decrease in $A_{Th}$ is explained by the lattice distortion due to thermal expansion (photothermal effect), which causes a large shift in $E_{ITT}$ and a significant decrease in absorption due to interband transitions.

\section{\label{sec:level6}Summary}
Using a microscopic femtosecond optical pump-probe technique, we observed both the nonthermal (NT) and the thermalized (Th) electron signal amplitude from the Al thin film on a Si cantilever. We found that the Th electron signal amplitude $A_{Th}$ varies linearly with the probe position, and $A_{Th}$ shows significant drops at the cantilever free end. The unique position-dependent properties of samples were explained by the change in interband transition threshold (ITT) $\Delta E_{ITT}$ caused by the position-dependent temperature rise and the associated thermal expansion by light absorption.
In particular, the validity of our method is supported by the parallel band structure, which exhibits remarkable optical properties in Al, and the detection of small strains using extremely intense light, which can only be realized with a light source such as a femtosecond laser.

Furthermore, our method does not require the use of mechanical resonance of the cantilever, which is necessary for detecting radiation pressure and photothermal effects. We argue that this study provides powerful measurement methods in various cantilever application techniques, such as mechanical resonators \cite{Kleckner2006,Ma2015}, calorimetric sensors \cite{doi:10.1063/1.4795625,doi:10.1063/1.1144509}, and nanolithography \cite{doi:10.1063/1.1785860,Milner2008}. Additionally, our findings provide useful insight into methods for measuring and controlling the optical properties of thin films \cite{Rold_n_2015} by exploiting the highly sensitive mechanical response of cantilevers. Moreover, further microscopic femtosecond optical pump-probe experiments on materials and microstructures are expected to lead to advanced ultrafast scanning probe microscopy systems.

\begin{acknowledgments}
This work was supported by Grant-in-Aid for JSPS Fellows (Grant Number. 22J11423) and CREST, JST (Grant Number. JPMJCR1875), Japan.
\end{acknowledgments}

\appendix
\section{Fitting results of transient reflectivity}
\label{sec:Ap_A}

\begin{table}
\caption{\label{tab:rAlc_coef}Coefficients ($A_0$, $A_1$) and their the standard deviations of the fitting parameters as a function of sampling position $x$ by linear fitting ($A_0+A_1x$) for the time-domain signal of the $L$~($\mu$m)-length cantilever.}
\begin{ruledtabular}
\begin{tabular}{cccc}
 &$L (\mu \mathrm{m})$&$A_0$&$A_1$\\
\hline
&160&$(971\pm6)\times10^{-6}$&$(-3.03\pm0.06)\times10^{-6}$\\
$A_{NT}$&200&$(796\pm6)\times10^{-6}$&$(-2.44\pm0.06)\times10^{-6}$\\
&240&$(779\pm12)\times10^{-6}$&$(-2.72\pm0.09)\times10^{-6}$\\
\hline
&160&$36.4\pm0.50$&$(-3.24\pm0.54)\times10^{-3}$\\
$\tau_p$ (fs)&200&$36.1\pm0.50$&$(-4.74\pm0.44)\times10^{-3}$\\
&240&$37.2\pm0.53$&$(-2.17\pm0.39)\times10^{-3}$\\
\hline
&160&$23.8\pm0.10$&$(3.18\pm1.09)\times10^{-3}$\\
$\tau_{NT}$ (fs)&200&$22.9\pm0.08$&$(8.75\pm0.69)\times10^{-3}$\\
&240&$21.7\pm0.17$&$(15.8\pm1.22)\times10^{-3}$\\
\hline
&160&$(164\pm1.3)\times10^{-6}$&$(-0.358\pm0.015)\times10^{-6}$\\
$A_{Th}$&200&$(157\pm1.1)\times10^{-6}$&$(-0.378\pm0.010)\times10^{-6}$\\
&240&$(162\pm1.3)\times10^{-6}$&$(-0.492\pm0.009)\times10^{-6}$\\
\hline
&160&$205\pm1$&$(154\pm15)\times10^{-3}$\\
$\tau_{Th}$ (fs)&200&$198\pm2$&$(273\pm14)\times10^{-3}$\\
&240&$195\pm1$&$(316\pm11)\times10^{-3}$\\
\hline
&160&$768\pm7$&$(548\pm80)\times10^{-3}$\\
$\tau_{ep}$ (fs)&200&$792\pm3$&$(643\pm29)\times10^{-3}$\\
&240&$767\pm5$&$(438\pm39)\times10^{-3}$\\
\end{tabular}
\end{ruledtabular}
\end{table}

The fitting parameters $A_{Th}$, $A_{NT}$, $\tau_p$, $\tau_{NT}$, $\tau_{Th}$ and  $\tau_{ep}$ were successfully extracted using Eq.~(\ref{eq:fit_func}). 
All coefficients used for the analysis are summarized in Table~II. 
Here, $A_0$ is the value at the intercept of $x=0$ (the root of the cantilever), and 
is comparable to the value obtained from Al film on a stable substrate (see Section~\ref{subsec:level3a}). $A_1$ is the slope of the line associated with the magnitude of the position dependency of the corresponding parameter. Although we discuss about  $A_{Th}$ in the main text, other parameters $A_{NT}$, $\tau_p$, $\tau_{NT}$, $\tau_{Th}$ and $\tau_{ep}$ also indicate some position dependency, for which it is required to perform more experimental and theoretical analyses in the near future.

\section{Derivation of equations for radiation pressure and strain for rectangular-section cantilever}
\label{sec:Ap_B}

In this section, we present the expression for the force $F$ due to radiation pressure and the derivation of equations for the strain and the deviation on the cantilever. The origin of the radiation pressure is the momentum of the pulsed photon given by $\bm{p}=(E_i/c^2)\bm{v}$, where $E_i$ is the energy of the incident light pulse, $\bm{v}$ and $c$ are the photon velocity and speed ($|\bm{v}|=c$), respectively \cite{Grant,Inan}.
From this equation, the momentum changes in the direction of the $z$-axis when the light is incident on a medium with reflectivity $R$ at the incidence angle $\theta$ is $\Delta p=(1+R)\cos{\theta}\cdot(E_i/c)$. Since $F$ is $\Delta p$ divided by $\tau_p$, and $E_i$ is the time-averaged power $P$ divided by $f_{rep}$, the above equation can be rewritten as,
\begin{equation}
\label{eq:F_RP}
F=(1+R)\cos{\theta}\cdot\frac{P}{c}\frac{1}{\tau_p\cdot f_{rep}}.
\end{equation}
Eq.~(\ref{eq:F_RP}) is also derived by using the relations where the absorptivity $A$ is $A=1-R$ in a medium with negligibly small transmittance, and the radiation pressure is expressed by the sum of the absorbed light component ($\propto A$) and the reflected light component ($\propto 2R$) as $F= (2R+A)P/c$ \cite{Ma2015}.
Using Eq.~(\ref{eq:F_RP}), $F$ is calculated as 274~$\mu$N given our experimental condition ($R=0.814$, $\theta=45^{\circ}$, $P=154$~mW, $\tau_p=30$~fs and $f_{rep}=80$~MHz). The value of $F$ was used in the main text. 
On the other hand, since the cantilever response ($f_{0}$: 73 - 280~kHz) is much slower than the repetition rate of laser pulses (80~MHz), this force is time averaged. The time-averaged force due to radiation pressure in the steady-state cantilever deflection is about only 0.7 nN calculated by Eq.~(\ref{eq:F_RP}) for $\tau_p\cdot f_{rep}=1$ (i.e., corresponding to a cw light at 154 mW).

\begin{figure}
\includegraphics[width=\linewidth]{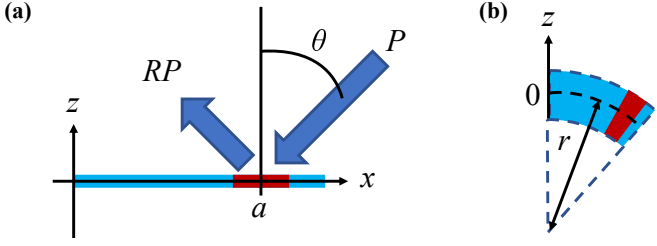}
\caption{\label{fig:Appendix}(a) Description of the force from radiation pressure by pulse irradiation and (b) consequent bend upon a cantilever (blue line). The irradiated positions are marked as red on the blue line. 
}
\end{figure}

Next, we take the $z$-axis with the center of the cantilever in the thickness direction as the origin and the surface side as positive [displayed in Fig.~\ref{fig:Appendix}(a)].
Here, the strain $e$ as a function of $z$ is $e(z)=z/r$, where $r$ is the curvature of the cantilever [illustrated in Fig.~\ref{fig:Appendix}(b)].
Based on kinetic analysis for rectangular section cantilever \cite{Sarid,s7091757}, $r$ is given by
\begin{equation}
\label{eq:kinetics1}
\frac{1}{r}=\frac{M(x)}{YI},
\end{equation}
where $M(x)$ is the bending moment, $Y$ is Young's modulus, and $I$ the area moment of inertia. 
When the light pressure occurs at position $x=a$, the bending moment $M(x)$ is given by
\begin{equation}
\label{eq:kinetics2}
M(x)=
	\begin{cases}
	F(a-x) & \text{for $x< a$,}\\
	0 & \text{for $x\geq a$.}\\
	\end{cases}
\end{equation}

For rectangular-section cantilever, $I$ is derived as,
\begin{equation}
\label{eq:kinetics3}
I=\frac{wd^3}{12}.
\end{equation}
where $w$ and $d$ are the width and thickness of the cantilever, respectively. By solving for $e(z)$ using the above equations, we obtain the following equation:
\begin{equation}
\label{eq:kinetics4}
e(z)=
\begin{cases}
	\displaystyle{\frac{12F(a-x)}{Ywd^{3}}z} & \text{for $x< a$,}\\
	0 & \text{for $x\geq a$.}\\
	\end{cases}
\end{equation}
Here, $x=a$ represents that the pump and probe beams are spatially overlapping. Except $x<a$, the $e(z)$ will becomes zero, which means that no observable static strain as discussed in previous sections.

Furthermore, expressing the deformation of the cantilever using the deviation in the $z$ direction $\delta_{z}(x)$ as a function of $x$, in the case of small curvature $r$ of the cantilever, 
\begin{equation}
\label{eq:kinetics5}
\frac{1}{r}=\frac{\partial^{2}}{\partial x^{2}}\delta_{z}(x)
\end{equation}
is valid. Equations~(\ref{eq:kinetics1}) and (\ref{eq:kinetics5}) allow us to formulate a differential equation for $x$. Under the conditions of Eq.~(\ref{eq:kinetics2}), the expression for $\delta_{z}(x)$ can be written as in the following equations,
\begin{equation}
\label{eq:kinetics6}
\delta_{z}(x)=
	\begin{cases}
	\displaystyle{\frac{2(3a-x)x^{2}}{Ywd^{3}}}F & \text{for $x< a$,}\\
	\displaystyle{\frac{2(3x-a)a^{2}}{Ywd^{3}}}F & \text{for $x\geq a$.}\\
	\end{cases}
\end{equation}
For $x=a$, the equation yields $\delta_{z}(x)=(4a^3F)/(Ywd^3)$, and the deviation is maximum when irradiated at the cantilever free end ($a=L$). Here, using the following equation for the spring constant $k_{z}$ of the cantilever,
\begin{equation}
\label{eq:kinetics7}
k_{z}=\frac{Ywd^3}{4L^3},
\end{equation}
the maximum deviation can be calculated using $\delta_{z}(L)=F/k_{z}$.
Therefore, the maximum deviation realized by the time-averaged radiation pressure in this experiment is 0.3~nm for $k_{z}=2.2$~N/m for the 240~$\mu$m-length cantilever, and we believe that the effect of deformation due to the time-averaged radiation pressure at any position of the cantilever is negligibly small.

\section{Derivation of equations for temperature rise on rectangular-section cantilever}
\label{sec:Ap_C}
Since the heat conduction of cantilever depends on its material and shape, to obtain an accurate temperature distribution, it is necessary to solve the three-dimensional diffusion equation by the finite element method \cite{https://doi.org/10.1049/mnl.2013.0352,Milner2010}. On the other hand, if the length $L$ of the cantilever is sufficiently larger than its width $w$ and thickness $d$, the temperature $T$ is assumed to be uniform over the entire cross-section, and the temperature distribution is described by the one-dimensional thermal diffusion equation as,
\begin{equation}
\label{eq:1D_dif_eq1}
\rho C_{l}\frac{\partial T}{\partial t}-\frac{\partial}{\partial x}\left[\lambda(T)\frac{\partial T}{\partial x}\right]=q(x),
\end{equation}
where $\rho$ is the mass density, $\lambda$ is the thermal conductivity at the temperature $T$, and $q$ is the heat source/sink density at the position $x$ \cite{WOS:000649074000009}.
In our measurements, the characteristic time of heat conduction $\rho C_{l}L^{2}/\lambda$ is on the scale of sub miliseconds, which is much longer than the repetition period of the pulsed light ($12.5$~ns) and much shorter than the integration time ($>100$~s). Therefore, the temperature distribution generated by repeated pulse irradiation can be regarded as stationary ($dT/dt=0$) \cite{WOS:000649074000009}.
Furthermore, we assume that the heat transfer to the ambient environment is negligibly small compared to the heat conduction through the cantilever. The surfaces of cantilever are thermally insulated except for the heating point $a$, then Eq.~(\ref{eq:1D_dif_eq1}) is transformed as follows
\begin{equation}
\label{eq:1D_dif_eq2}
\frac{\partial}{\partial x}\left[\lambda(T)\frac{\partial T}{\partial x}\right]=\frac{(1-R)P}{wd}\delta_{D}(x-a),
\end{equation}
where $\delta_{D}$ is the Dirac's distribution. The distribution of temperature rise $\Delta T_{l}(x)=T_{l}(x)-T_{0}$ can be obtained by integrating Eq.~(\ref{eq:1D_dif_eq2}) twice under the boundary conditions assuming isothermally clamped ends [$T_{l}(x=0)=T_{0}$] and thermally insulated free ends ($[dT_{l}/dx]_{x=L}=0$), as
\begin{equation}
\label{eq:1D_dif_eq_sol1}
\begin{cases}
	\displaystyle{\int^{T_{0}+\Delta T_{l}}_{T_{0}}\lambda(T')dT'=\frac{(1-R)P}{wd}x} & \text{for $x\leq a$,}\\
	\Delta T_{l}(x)=\Delta T_{l}(a) & \text{for $x> a$.}\\
	\end{cases}
\end{equation}
If $\lambda(T)$ is a constant as $\lambda(T) = \lambda_{0}$, the integral is solved as
\begin{equation}
\label{eq:1D_dif_eq_sol2}
\Delta T_{l}(x)=\frac{(1-R)P}{w\lambda_{0} d}x
\end{equation}
for $x\leq a$ \cite{WOS:000649074000009}. This equation can be extended to an expression applicable to a bi-layered cantilever \cite{doi:10.1063/1.4795625,doi:10.1063/1.1144509,doi:10.1063/1.2829999,s7091757} by replacing $\lambda_0$ and $d$ with those of the respective layers, and the equation adapted to our experimental conditions is obtained as in Eq.~(\ref{eq:1D_dif_eq_sol3}).
This temperature rise causes significant thermal expansion, but does not exceed the melting point of Al ($\sim$933~K). The temperature dependence of $\lambda$ leads the temperature distribution to be nonlinear \cite{WOS:000649074000009}, but our experimental results did not show a nonlinear trend because the rised temperature ($300+474$~K) was smaller than the melting point of Al. Other factors involved in modifying the model (e.g., boundary condition issues) are beyond the scope of this paper.

\nocite{*}
\bibliography{Ichikawa_2023Jan30}

\end{document}